\documentclass[prd,floats,aps,nofootinbib,axodraw,amssymb,preprintnumbers]{revtex4}
\usepackage{amsmath,amsfonts}
\usepackage{subfigure}
\usepackage{verbatim}
\usepackage{slashed}
\usepackage{feynmf}
\usepackage{graphicx}
\usepackage{epstopdf}
\def\beq{\begin{equation}}
\def\eeq{\end{equation}}
\def\bea{\begin{eqnarray}}
\def\eea{\end{eqnarray}}

\def\leqn#1{(\ref{#1})}

\def\pslash{\not{\hbox{\kern-4pt $p$}}}
\def\qslash{\not{\hbox{\kern-4pt $q$}}}
\def\lv{\not{\hbox{\kern-4pt $L$}}}
\def\lsim{\mathrel{\raise.3ex\hbox{$<$\kern-.75em\lower1ex\hbox{$\sim$}}}}
\def\gsim{\mathrel{\raise.3ex\hbox{$>$\kern-.75em\lower1ex\hbox{$\sim$}}}}
\def\ifmath#1{\relax\ifmmode #1\else $#1$\fi}

\usepackage{graphicx}
\usepackage{bm}
\preprint{SLAC-PUB-16173}
\preprint{MITP/14-103}
\preprint{SCIPP 14/17}

\begin{document}

\title{Perturbative Unitarity Constraints on Gauge Portals}
\bigskip
\author{Sonia El Hedri$^{a,b}$, William~Shepherd$^c$  and Devin~G.~E.~Walker$^b$}
\address{$^a$Institut f\"ur Physik (THEP) Johannes Gutenberg-Universit\"at,
D-55099, Mainz, Germany}
\address{$^b$ SLAC National Accelerator Laboratory, 2575 Sand Hill Road, Menlo Park, CA 94025, U.S.A.}
\address{$^c$ Santa Cruz Institute for Particle Physics and Department of Physics, Santa Cruz, CA 95064, U.S.A.}

\begin{abstract}
\noindent
Dark matter that was once in thermal equilibrium with the Standard Model is generally prohibited from obtaining all of its mass from the electroweak phase transition.  This implies a new scale of physics and mediator particles to facilitate dark matter annihilation.  In this work, we focus on dark matter that annihilates through a generic gauge boson portal.  We show how partial wave unitarity places upper bounds on the dark gauge boson, dark Higgs and dark matter masses.  Outside of well-defined fine-tuned regions, we find an upper bound of 9 TeV for the dark matter mass when the dark Higgs and dark gauge bosons both facilitate the dark matter annihilations.  In this scenario, the upper bound on the dark Higgs and dark gauge boson masses are 10 TeV and 16 TeV, respectively.  When only the dark gauge boson facilitates dark matter annihilations, we find an upper bound of 3 TeV and 6 TeV for the dark matter and dark gauge boson, respectively.  Overall, using the gauge portal as a template, we describe a method to not only place upper bounds on the dark matter mass but also on the new particles with Standard Model quantum numbers.  We briefly discuss the reach of future accelerator, direct and indirect detection experiments for this class of models.
\end{abstract}

\maketitle

\section{Introduction}
\label{sec:intro}
\noindent
Understanding the nature of dark matter (DM) is one of the most pressing unresolved problems in particle physics.  Dark matter is needed to understand structure formation, the observed galactic rotation curves~\cite{Kowalski:2008ez,Ahn:2013gms,Beringer:1900zz} and the acoustic peaks in the cosmic microwave background~\cite{Ade:2013lta}.  Moreover, the dark matter relic abundance is measured to be~\cite{Ade:2013lta}  
\begin{equation}
h^2\, \Omega_c = 0.1199 \pm 0.0027.  \label{eq:relic}
\end{equation}
A compelling argument for the origin of this abundance is to assume dark matter was once in thermal contact with the baryon-photon plasma during the early universe.  Since all known forms of matter in the universe were once in thermal equilibrium, this type of dark matter is theoretically persuasive.  In this scenario, the measured relic abundance is controlled by dark matter annihilations into Standard Model (SM) particles.  Because of constraints from the observed large scale structure in the universe, dark matter must be stable and non-relativistic (cold) when departing thermal equilibrium~\cite{Ahn:2013gms}.  
\newline
\newline
The Standard Model (SM) alone cannot account for the missing matter in the universe~\cite{Bertone:2004pz}.  Current experimental constraints, however, provide some guidance on the structure of the underlying theory.  For example, the lack of large missing energy signatures at the Large Hadron Collider (LHC)~\cite{Aad:2014tda,ATLAS:2014wra,Aad:2013oja,Aad:2014vka,Aad:2014kra,Chatrchyan:2014lfa,Khachatryan:2014tva,Khachatryan:2014rwa,Khachatryan:2014rra,Chatrchyan:2013lya,Chatrchyan:2013xna} and other colliders~\cite{Aaltonen:2012jb,Aaltonen:2013har,CDF:2011ah,Abachi:1996dc,Abazov:2012qka,Abdallah:2003np,Acciarri:1997dq,Abbiendi:2004gf} suggest that dark matter is either heavy or has very small couplings with the SM so that it is not produced in high-energy collisions.  Additionally, direct detection experiments~\cite{Aprile:2012nq,Ahmed:2011gh,Akerib:2013tjd}, updated precision electroweak constraints, and precision Z-pole experiments~\cite{devinjoannetim,Baak:2012kk,ALEPH:2005ab} all severely constrain the direct coupling of dark matter to the SM Higgs and/or Z~bosons.  These constraints all imply dark matter cannot obtain all of its mass from the SM Higgs alone~\cite{devinjoannetim}.  Thus, if dark matter is a weakly interacting massive particle (WIMP), we are led to scenarios where new mediators facilitate dark matter interactions with the SM.  Moreover, a new fundamental scale of physics is needed that is (at least partly) responsible for the dark matter mass.  Mediator-facilitated interactions help to evade current experimental constraints by partially decoupling the dark matter from the SM.  Should these scenarios be realized in nature, the discovery of the mediator particles would be an important step in understanding the nature of dark matter.  It is therefore crucial to place bounds on the masses and couplings of these mediators.  The most popular ways for dark matter to annihilate via a mediator particle are through the Higgs~\cite{Patt:2006fw} boson, through scalars that are colored or charged, or via a new neutral gauge boson.  Some of us considered the perturbative unitarity constraints on the Higgs portal in~\cite{Walker:2013hka,Betre:2014sra,Betre:2014fva}.  In this work, we focus on placing constraints on a scenario where fermionic dark matter is charged under a new, dark gauge group, $U(1)_D$.  This gauge group is spontaneously broken by a dark Higgs, $\Phi$, generating a massive, dark $Z'$ boson.  This boson is also known in the literature as a dark photon.  The dark $Z'$ mixes kinetically as well as through mixed mass terms with the SM $Z$ boson.  %associated with SM hypercharge.  %In addition to the dark matter and dark gauge boson $Z'$, a new dark Higgs $H$ is present that breaks $U(1)_D$ by getting a vev and therefore facilitates the spontaneous symmetry breaking. 
Thus, the mixing between the hidden sector and the SM allows dark matter (DM) to annihilate via the Higgses, $Z$ and $Z'$ bosons. 
\newline
\newline
We apply unitarity constraints in a manner reminiscent of Griest and Kamiokowski~\cite{Griest:1989wd}.  However, there are important differences:  Here we focus on perturbative unitarity constraints which determine, in particular, when the dark matter couplings become strong.  WIMP dark matter and perturbativity have always had an important conceptual association. Dark matter masses that violate the perturbative unitarity bounds imply the dark matter is efficiently forming bound states as well as annihilating as the temperature decreases toward the thermal decoupling temperature. Because the dark matter annihilates into lighter states, the annihilation diagrams can be altered (and sometimes dressed with these lighter states) to produce diagrams in which the bound states decay. The dark matter decays have a lifetime well shorter than the age of the universe.  Thus, %
%In all, we surmise that 
dark matter with a mass beyond the perturbative unitarity bounds is not an asymptotic state; this leads to a scenario without a viable dark matter candidate. Note, it is well known that viable dark matter candidates exist that are the result of strongly coupled or confining hidden sectors. However, in these models the dark matter annihilation processes are still perturbative~\cite{Shepherd:2009sa}.  We show our perturbative unitarity constraints are improved in comparison to the updated Griest and Kamiokowski bounds~\cite{Profumo:2013yn}.  Of central importance is the fact that our methodology places constraints on any particle associated with the dark matter annihilation.  For this paper, our bounds on the masses and couplings of the new Higgs and dark gauge boson are novel.
\newline
\newline
Our basic perturbative unitarity arguments are straightforward.  The DM annihilation cross section depends on the masses of the dark matter, the dark Higgs, and the $Z'$, as well as the dark matter couplings to the dark Higgs and dark $Z'$.  As the dark matter gets heavier, its annihilation cross section decreases. In order for heavy dark matter to satisfy the relic abundance constraints, it must annihilate more efficiently and therefore have sizable couplings to the SM and hidden sectors.  Eventually, the couplings required to obtain the correct relic abundance are so large that perturbative unitarity is violated. Perturbative unitarity arguments therefore set an upper bound on the dark matter mass. If the dark Higgs and gauge boson masses are raised to be larger than the dark matter mass, fewer (and more suppressed) annihilation channels are available.  The annihilation cross section in these regimes of parameter space is thus diminished.  Therefore these arguments yield bounds on the dark matter mass as well as on the mass of any other particle involved with the dark matter annihilation.  %Therefore, any upper bound on the DM mass corresponds to neighboring upper bounds on the masses of the other hidden sector particles.
\newline
\newline
In the next section, we  introduce a generic $U(1)_D$ model on which to place our unitary bounds and introduce the parameters that need to be constrained as well as the constraints from electroweak precision tests (EWPT).  In Section III., we show how to apply unitarity constraints on the various sectors of the model.  Section IV.~details how relic abundance and direct detection constraints on the DM sector impact the masses and couplings of the theory.  Section V.~gives our results by detailing the bounds on the particle masses obtained by applying the EWPT, unitarity and relic abundance constraints.  Conclusion and Appendices follow.

\section{A Representative Model}
\label{Sec: sec3}
\noindent
We extend the SM with an additional $U(1)_D$ gauge group with coupling $g_D$ that is spontaneously broken at a high scale.  This dark group is associated to a dark gauge boson that mixes kinetically and via mass terms with SM hypercharge.  The $U(1)_D$ gauge group is broken by a new, dark Higgs, that gets a vev $u$. The model then includes two Higgs fields
\begin{align}
%\phi &= \begin{pmatrix}
%G^\pm\\
%\frac{v + h+ i G^0}{\sqrt{2}}
%\end{pmatrix}\quad\quad\Phi = \frac{u + H + i G'^{\,0}}{\sqrt{2}}.
%%%
H = {1 \over \sqrt{2}}\,\begin{pmatrix}
\sqrt{2}\,G^\pm\\
 v + h + i \,G^0
\end{pmatrix}
&& \Phi = {1 \over \sqrt{2}}\bigl(u + \rho + i \,G_\rho^0\bigr).
\end{align}
where $H$ is the SM Higgs.  We also introduce a DM candidate $\chi$, which is a chiral fermion, neutral under the SM gauge groups but charged under $U(1)_D$.  %Before diagonalizing the kinetic terms, 
All SM particles are taken to be neutral under $U(1)_D$. The dark charge assignments for the DM and the dark Higgs are
\begin{align}
Q_\Phi = -2 \qquad \qquad Q_{\chi_L} = -1 \qquad \qquad Q_{\chi_R} = 1.
\end{align}
Anomaly cancellation mandates the introduction of the second chiral fermion. In this work, we take this additional fermion to be much heavier than the other particles so that it does not have any influence on the final results.
\newline
\newline
We adopt the notations and conventions from~\cite{Babu:1997st}. The relevant parts of the lagrangian associated to new physics is
\begin{align}
%\mathcal{L} &= \mathcal{L}_\mathrm{kin}^\mathrm{gauge} + \mathcal{L}_\mathrm{kin}^\mathrm{DM}+ \mathcal{L}_\mathrm{yuk}^\mathrm{DM}+ \mathcal{L}_\mathrm{kin}^\mathrm{Higgs} + V(H,\Phi )
\mathcal{L} &= \mathcal{L}_\mathrm{gauge} + \mathcal{L}_\mathrm{DM} + \mathcal{L}_\mathrm{Higgs} 
\end{align}
where the dark matter and gauge sectors are
\begin{align}
% \mathcal{L}_\mathrm{kin}^\mathrm{gauge} &= -\frac{1}{4}\hat{B}_{\mu\nu}\hat{B}^{\mu\nu}-\frac{1}{4}\hat{Z'}_{\mu\nu}\hat{Z'}^{\mu\nu} + \frac{\sin\delta}{2}\hat{B}_{\mu\nu}\hat{Z'}^{\mu\nu}\\
% \mathcal{L}_\mathrm{kin}^\mathrm{DM} &= \bar \chi_L \slashed{D}_\mu \chi_L + \bar chi_R \slashed{D}_\mu \chi_R\\
% \mathcal{L}_\mathrm{Yuk}^\mathrm{DM} &= \lambda_\chi \bar\chi_L \Phi \chi_R + h.c.\\
% \mathcal{L}_\mathrm{kin}^\mathrm{Higgs} &= \left|D_\mu \phi\right|^2 + \left|D_\mu \Phi\right|^2\\
%V(\phi, \Phi) &= \lambda_1 \left( H^\dagger H - \frac{v^2}{2}\right)^2 + \lambda_2 \left( \Phi^\dagger\Phi - \frac{v^2}{2}\right)^2 + \lambda_3\left( H^\dagger H - \frac{v^2}{2}\right)\left( \Phi^\dagger\Phi - \frac{v^2}{2}\right) .
%
 \mathcal{L}_\mathrm{gauge} &= -\frac{1}{4}\hat{B}_{\mu\nu}\hat{B}^{\mu\nu}-\frac{1}{4}\hat{Z'}_{\mu\nu}\hat{Z'}^{\mu\nu} - \frac{\sin\delta}{2}\hat{B}_{\mu\nu}\hat{Z'}^{\mu\nu}\\
 \mathcal{L}_\mathrm{DM} &= \bar \chi_L \slashed{D}_\mu \chi_L + \bar \chi_R \slashed{D}_\mu \chi_R -  \lambda_\chi \bar\chi_L \Phi \chi_R + h.c.
 \end{align}
 and 
\begin{align}
 \mathcal{L}_\mathrm{Higgs} &= \left|D_\mu H\right|^2 + \left|D_\mu \Phi\right|^2 - V(H,\Phi ) \\
V(H, \Phi) &= \lambda_1 \left( H^\dagger H - \frac{v^2}{2}\right)^2 + \lambda_2 \left( \Phi^\dagger\Phi - \frac{u^2}{2}\right)^2 + \lambda_3\left( H^\dagger H - \frac{v^2}{2}\right)\left( \Phi^\dagger\Phi - \frac{u^2}{2}\right) .
\end{align}
is the Higgs sector.  The kinetic mixing is parameterized by the mixing angle $\delta$.  The kinetic terms can be diagonalized by defining new fields $B_\mu$ and $Z'_\mu$ such that~\cite{Babu:1997st}
\begin{align}
\begin{pmatrix}
\hat{B}_\mu \\
\hat{Z'}_\mu
\end{pmatrix} = 
\begin{pmatrix}
1 & -\tan\delta\\
0 & \sec\delta %\frac{1}{\cos\delta}
\end{pmatrix}
\begin{pmatrix}
B_\mu \\
Z'_\mu
\end{pmatrix}
\end{align}
where the hatted fields are the fields before diagonalizing kinetic mixing.  Denoting $g_1$ and $g_2$ as the SM hypercharge and weak couplings respectively, the covariant derivatives for the Higgs and DM fields then become
\begin{align}
D H &= \partial H -i g_2 W^a \sigma^a H - \frac{i g_1}{2} B H + \frac{i g_1}{2} \tan\delta\, Z' H\\
D \Phi &= \partial \Phi - \frac{2 i g_D}{\cos\delta} Z' \Phi\\
D \chi_R &= \partial \chi_R - \frac{i g_D}{\cos\delta} Z' \chi_R\\
D \chi_L &= \partial \chi_L + \frac{i g_D}{\cos\delta} Z' \chi_L.
\end{align}
Any SM particle with non-zero hypercharge will then acquire a dark charge.  We now have the effective dark gauge coupling
\begin{align}
g' = \frac{g_D}{\cos\chi}.
\end{align}
After symmetry breaking, the gauge bosons acquire masses. Kinetic mixing between $\hat{B}_\mu$ and $\hat{B'}_\mu$ induces a mass mixing between $Z$ and $Z'$. The mass mixing angle, $\xi$, is such that 
\begin{align}
\tan 2\xi = \frac{-2 \cos\delta\sin\delta \sin\hat\theta_W\hat{M}_{Z}^2}{\hat{M}_{Z'}^2 - \hat{M}_{Z}^2 \cos^2\delta + \hat{M}_{Z}^2 \sin^2\hat\theta_W\sin^2\delta}
\end{align}
where the hatted Weinberg angle $\hat{\theta}_W$ is such that 
\begin{align}
A_\mu &= \cos\hat\theta_W B_\mu + \sin\hat\theta_WW^3_\mu \qquad \qquad Z_\mu = -\sin\hat\theta_W B_\mu + \cos\hat\theta_WW^3_\mu
\end{align}
and the masses terms are
\begin{align}
\hat{M}_Z &= \frac{v}{2}\sqrt{g_1^2 + g_2^2} \quad\quad \hat{M}_{Z'} = 2 g' u.
\end{align}
EWSB also induces a mass mixing between the scalars $h$ and $\rho$. The mixing angle $\theta$ is given by
\begin{align}
\tan 2\theta = \frac{2\lambda_3\, u\, v}{\lambda_2 u^2 - \lambda_1 v^2}.
\end{align}
We denote the light and heavy mass eigenstates in the Higgs and gauge sector by $h_{1}$, $h_2$, $Z_1$ and $Z_2$ respectively. The lightest mass eigenstates, $h_1$ and $Z_1$ can be identified with the $125$ GeV SM Higgs and the SM $Z$ boson, respectively.
\newline
\newline
The DM acquires a mass through the Yukawa coupling to the dark Higgs
\begin{align}
m_\chi = \frac{\lambda_\chi}{\sqrt{2}}u.
\end{align}
Due to the charge assignements, the coupling of the DM to the dark gauge boson interaction eigenstate $Z'_\mu$ is purely axial-vector. The DM-Higgs Yukawa coupling also induces a purely scalar coupling between DM and the dark Higgs interaction eigenstate $\rho$. The DM interaction Lagrangian can then be written as
\begin{align}
\mathcal{L}_\mathrm{kin}^\mathrm{DM} \supset g'\bar\chi \gamma^\mu \gamma_5 \chi Z'_\mu - \frac{\lambda_\chi}{\sqrt{2}} \,\Phi\bar\chi_L \chi_R + h.c.
\end{align}
The model now has seven independent parameters
\begin{align}
\{\lambda_1, \lambda_2, \lambda_3, \chi, g', \lambda_\chi, u\}
\label{Eq: parameters}
\end{align}
to be constrained.

\begin{figure}
\includegraphics[width=\linewidth,natwidth=161,natheight=138]{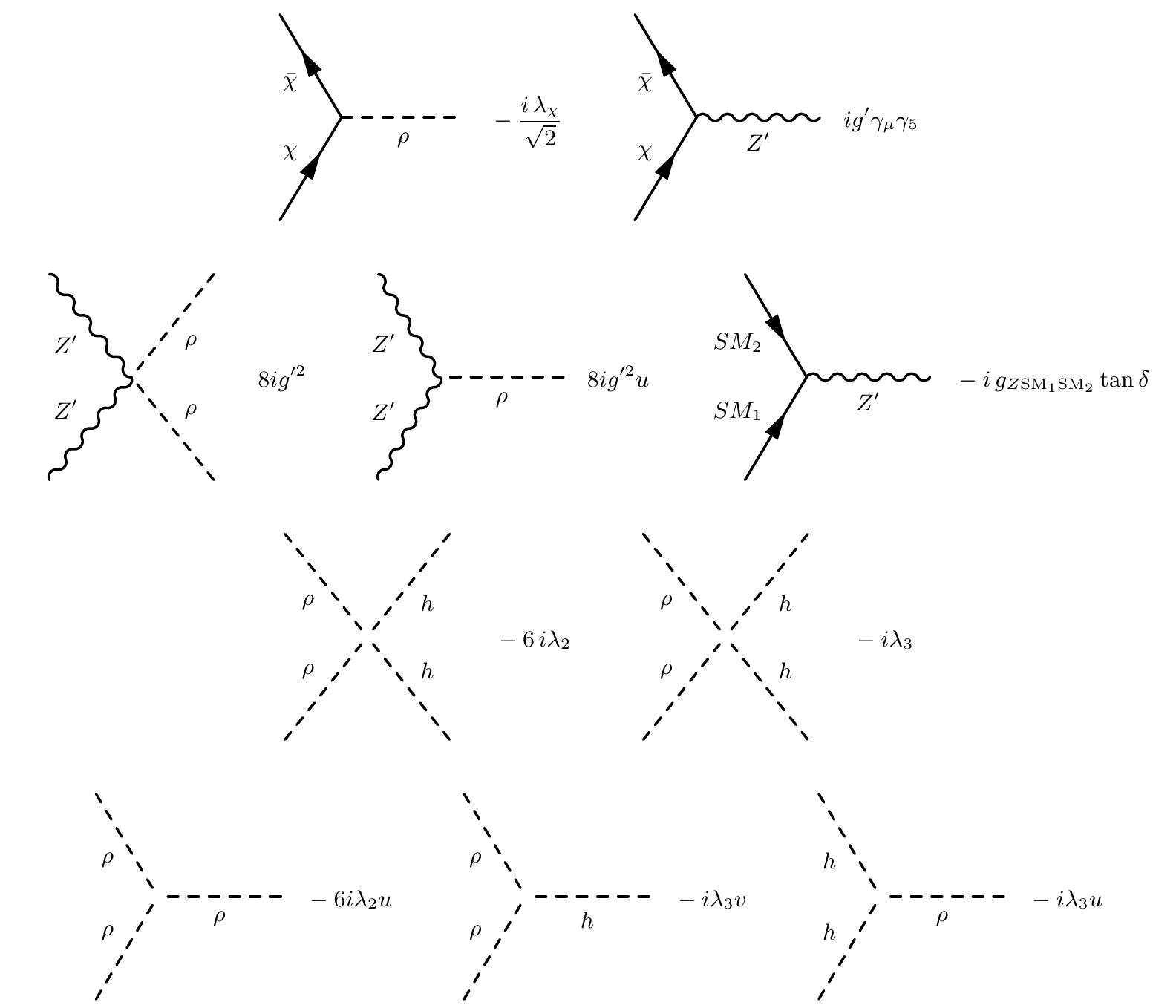}
\caption{\label{Fig: vertices} Vertices involving the hidden sector particles in the $U(1)_D$ model detailed in Sec.~\ref{Sec: sec3}}.
\end{figure}
\subsection{The Heavy Limit}
\noindent
Since we want to derive upper bounds on the masses of the particles in the hidden sector, we are particularly interested in the limit where these particles are heavy. In this limit, the mass mixings become negligible and we can work in the gauge eigenstate basis. Since kinetic mixing does not vanish at large masses, we work in the kinetic mixing eigenbasis, using the fields $h$, $\rho$, $Z$ and $Z'$. 
\newline 
\newline
The masses of the hidden sector particles in the heavy limit are
\begin{align}
m_\chi = \frac{\lambda_\chi}{\sqrt{2}} u \quad m_{\rho} = \sqrt{2\lambda_2}\,u \quad m_{Z'} = 2 g' u.
\end{align}
The kinetic mixing induces a non-zero coupling of the $Z'$ to SM particles of the form
\begin{align}
g_{Z' \mathrm{SM}_1\mathrm{SM}_2} &= \tan\delta\,\,g_{Z \,\mathrm{SM}_1\mathrm{SM}_2}. \label{eq:ffZprimecoupling}
\end{align}
This coupling does not vanish when the $Z'$ becomes heavy, and will play an important role in the s-channel annihilation of dark matter, as highlighted in Sec.~\ref{Sec: sec5}. The vertices involving hidden sector particles are shown in Fig.~\ref{Fig: vertices}. In this heavy limit, DM annihilation will be driven by the $\lambda_\chi$ and $g'$ couplings, and, in some channels, by the kinetic mixing $\sin\delta$.

\subsection{Precision Electroweak Constraints}
\noindent
Models with a hidden $Z'$ boson are generally expected to be constrained by precision electroweak measurements. Since the $Z$ pole mass, the electron charge and the physical Weinberg angle $\theta_W$ have been measured very precisely, we set them to be equal to their SM values. Keeping these parameters fixed allows other parameters to deviate from their SM values. In particular, $\hat \theta_W$ becomes
\begin{align}
\sin\hat\theta_W\cos\hat\theta_W &= \sin\theta_W \cos\theta_W \frac{M_{Z_1}}{ \hat{M}_Z},
\end{align}
which is then in general not equal to the physical Weinberg angle. As shown in \cite{Babu:1997st,Wells:2008xg,Peskin:1991sw,Holdom:1990xp}, the parameters that leads to the tightest precision electroweak constraints on the model introduced here are
\begin{align}
\Delta m_W &= (17\,\mathrm{MeV})\Upsilon\\
\Delta \Gamma_{l^+l^-} &= -(80\,\mathrm{keV})\Upsilon\\
\Delta\sin^2\theta_W^{eff} &= -(0.00033)\Upsilon
\end{align}
where
\begin{align}
\Upsilon &= \left(\frac{\tan\delta}{0.1}\right)^2\left(\frac{250\mathrm{GeV}}{m_{Z_2}}\right)^2.
\end{align}
We follow \cite{Wells:2008xg} and require $\Upsilon \le 1$.

\section{Constraints from Unitarity}
\label{Sec: sec4}
\noindent
This section details how to apply unitarity constraints on different scattering processes in the model studied. Although the scattering $S$ matrix is always unitary when resummed over all orders, it is not unitary at tree-level. As demonstrated in~\cite{Aydemir:2012nz}, unitarity is restored by loop corrections.  Schuessler and Zeppenfeld in~\cite{Schuessler:2007av,Schuessler:thesis} have derived a simple geometric argument that uses the tree-level scattering amplitudes to conservatively estimate the minimal amount of loop corrections needed to make a theory unitary. In particular, if the scattering amplitudes for a given process are too large at tree-level, large loop corrections are needed to restore unitarity and the theory is no longer perturbative. As mentioned in the introduction, if the couplings of the particle $\chi$ to Higgs or gauge bosons are non-perturbative, the DM candidate we have to consider is the $\chi \bar\chi$ bound state. The processes corresponding to the annihilation of $\chi$ and $\bar\chi$ will then correspond to decay modes for the true dark matter particle, which then cannot be a suitable DM candidate because it has fast decays. In order for our model to provide a viable explanation for DM, the couplings $\lambda_\chi$ and $g'$ thus must be perturbative.
\newline
\newline
In what follows, we apply the procedure outlined in~\cite{Schuessler:2007av,Schuessler:thesis} and explained in detail in the case of the NMSSM in~\cite{Betre:2014sra}. In order to enforce perturbativity, we require that the loop corrections to the scattering amplitudes studied have to be less than $40\%$ of the corresponding tree-level amplitudes. As shown in~\cite{Schuessler:2007av,Schuessler:thesis}, this perturbativity requirement translates into an upper bound on the eigenvalues of the partial-wave components of the transition operator $T$, defined as
\begin{align}
S = I + i\,T
\end{align}
with $S$, the S-matrix. Denoting the jth partial wave component of this operator by $\mathcal{T}^j$, the perturbativity requirement becomes
\begin{align}
\left|\mathcal{T}^j_{ii}\right| < \frac{1}{2}.
\end{align}
Generally, lower $j$ provide stronger bounds. In what follows, we will consider either the s-wave ($j = 0$) or the p-wave ($j = 1$) components of the scattering amplitudes. In what follows, we will work in the limit where
\begin{align}
s \gg m_{H_2}, m_{Z_2}, v. %\mathrm{EW\,scale}.
\end{align}

\subsection{Dark Matter Scattering Amplitudes}
\begin{figure}
\centering
\includegraphics[width=0.5\linewidth,natwidth=150,natheight=150]{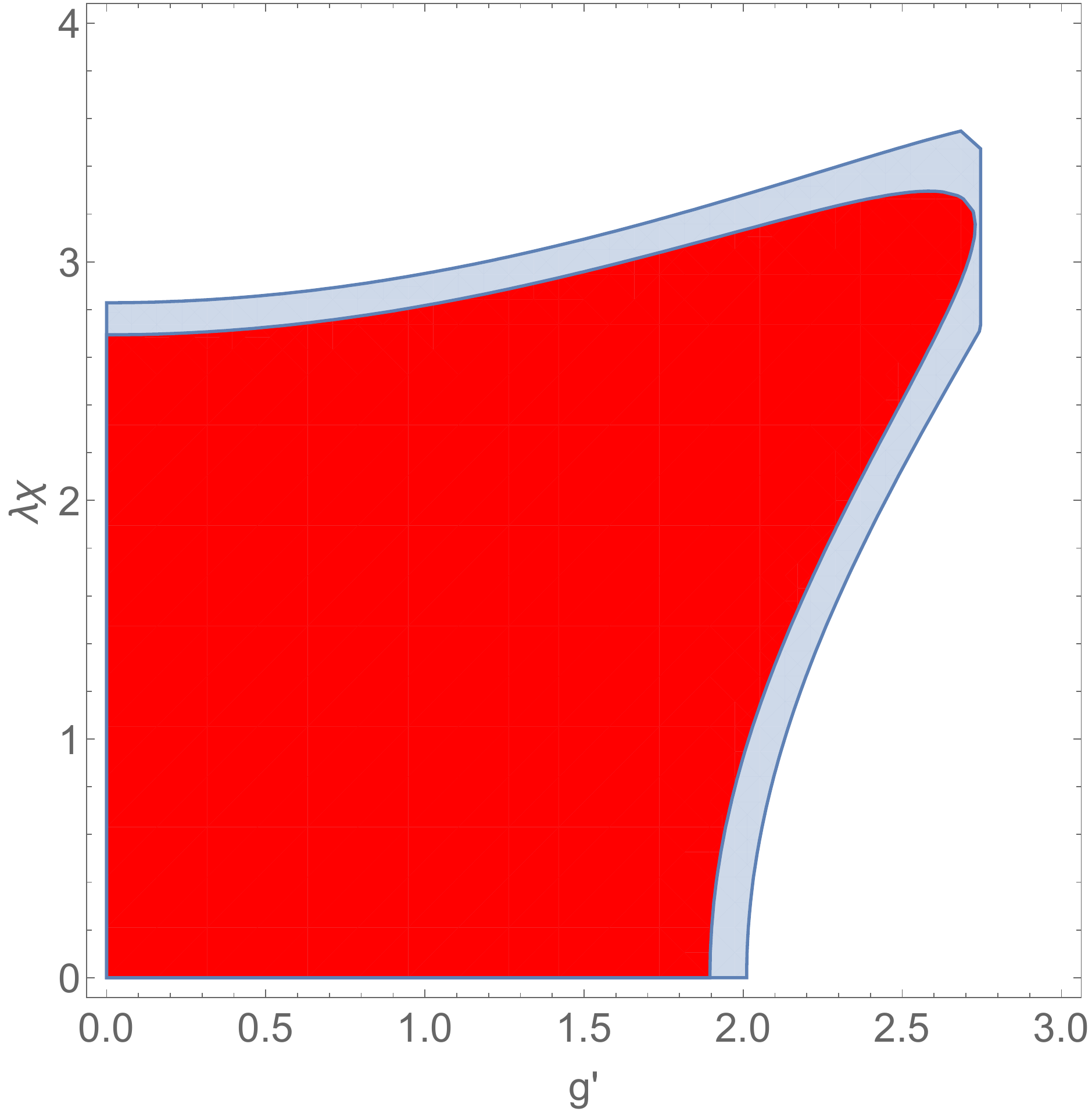}
\caption{\label{Fig: unitarityFull} Unitarity bounds on $g'$ and $\lambda_\chi$ for $\lambda_2 = 0$ (blue) and $\lambda_2 = 4$ (red). To enforce perturbativity, we require the loop corrections to the scattering amplitudes to be less than $40\%$ of the tree-level value, which corresponds to the bounds shown in Eq.~\ref{Eq: amp}.}
\end{figure}

\noindent
Ideally, setting bounds on scattering amplitudes in the s-wave provides the best constraints. However, in the case of fermion scatterings in the large $s$ limit, diagrams with an intermediate gauge boson exhibit a Coulomb-like pole. In this case, the amplitudes are logarithmically divergent at tree-level. Once higher order corrections are applied, we find that this logarithmic divergence corresponds in fact to the first term of the expansion of a phase and does not lead to perturbativity/unitarity violations. Applying unitarity bounds on tree-level scattering amplitudes then does not work in the s-wave, as it is sensitive to the logarithmic divergence. The p-wave scattering is not sensitive to this divergence, and unitarity constraints in the p-wave provide generally weaker but still non-trivial bounds. 
\newline
\newline
In the p-wave, we consider the following fermion and scalar pairs
\begin{align}
\bigl(\chi_+\bar{\chi}_+, \chi_-\bar{\chi}_-, \chi_+\bar{\chi}_-, \chi_-\bar{\chi}_+, \rho\, Z', \rho\, Z, h\,Z, h\,Z' \bigr).
\end{align}
The $+$ and $-$ are the fermion helicities (also right and left). Since we work in the large $s$ limit, we can work in the interaction eigenstate basis. The p-wave scattering matrix in this basis is
\begin{align}
    \mathcal{T} &= \frac{-1}{64\sqrt{2}}\begin{pmatrix}
       -6 g'^2 & 0 & 0 & 0 & 0 & 0 & 0 & 0\\
        0 & -6 g'^2 & 0 & 0 & 0 & 0 & 0 &0\\
        0 & 0 & 4 g'^2 & 2g'^2 & \frac{16}{3}g'^2 - 4 \lambda_\chi^2 & 0 & 0 & 0\\
        0 & 0 & 2 g'^2 & 4 g'^2 & -\frac{16}{3}g'^2 + 4 \lambda_\chi^2 & 0 & 0 & 0\\
        0 & 0 & \frac{16}{3} g'^2 - 4\lambda_\chi^2 & -\frac{16}{3}g'^2 + 4\lambda_\chi^2 & 2\lambda_2 + 5g'^2 & 0 & 0 & 0\\
        0 & 0 & 0 & 0 & 0 & \lambda_3 & 0 & 0\\
        0 & 0 & 0 & 0 & 0 & 0 & 2\lambda_1 & 0\\
        0 & 0 & 0 & 0 & 0 & 0 & 0 & \lambda_3
    \end{pmatrix}.
\label{Eq: FermionMatrix}
\end{align}
We enforce the perturbativity requirement by requiring that the eigenvalues of $\mathcal{T}$ verify
\begin{align}
 \left|\mathcal{T}_{ii}\right| < \frac{1}{2}.
\label{Eq: amp}
\end{align}
Fig.~\ref{Fig: unitarityFull} shows the resulting unitarity bounds on $\lambda_\chi$ and $g'$ for $\lambda_2 = 0$ and $\lambda_2 = 4$. The bounds depend only weakly on $\lambda_2$ and can be estimated as
\begin{align}
g' \lesssim 2.8 \qquad \qquad \qquad \lambda_\chi \lesssim 3.5.
\end{align}

\subsection{Scalar-Scalar scattering amplitudes}
\begin{figure}
\centering
\includegraphics[width=0.5\linewidth,natwidth=150,natheight=150]{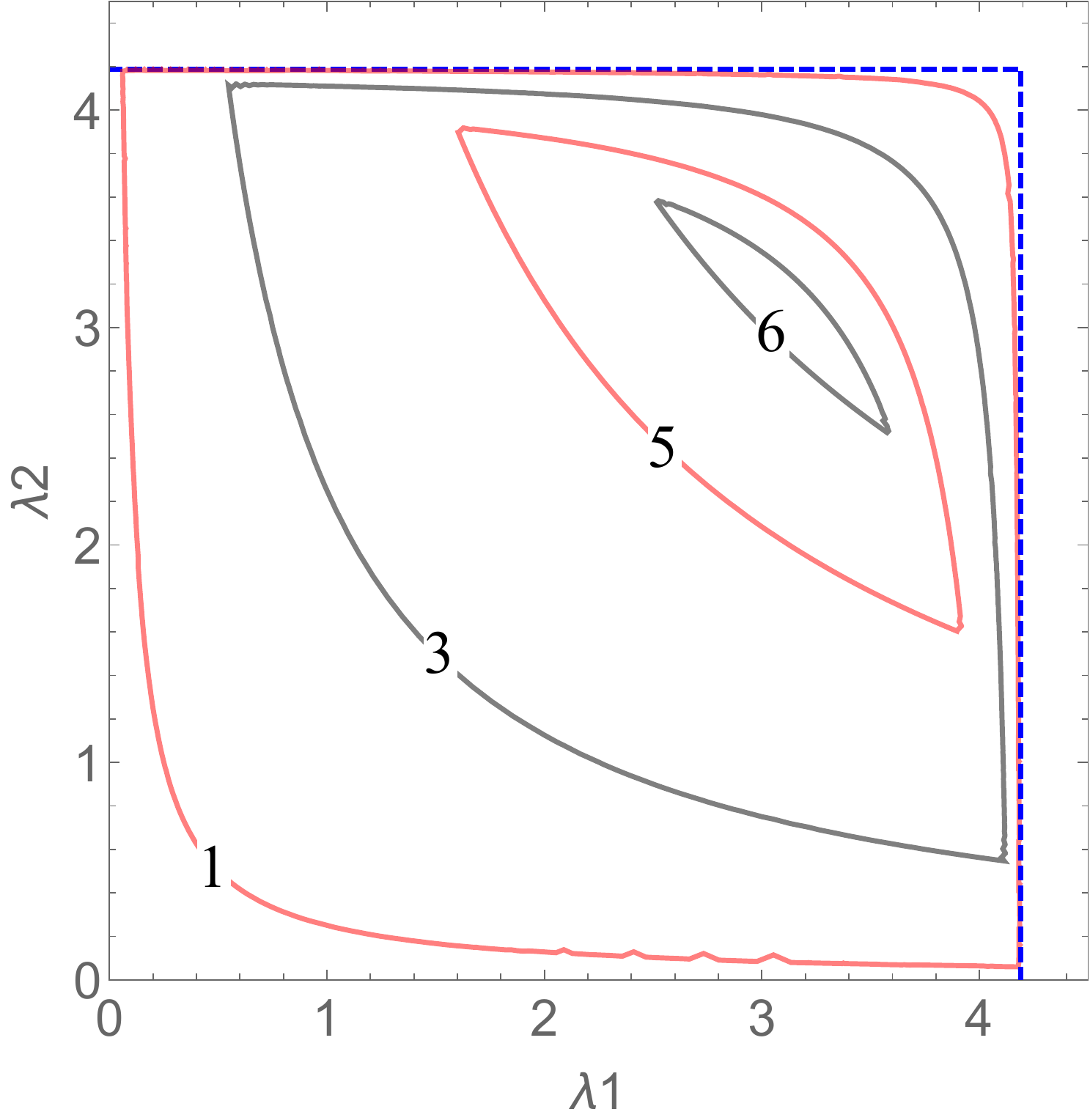}
\caption{\label{Fig: maxl3} Contours of the maximum values of $\lambda_3$ versus $\lambda_1$ and $\lambda_2$. We apply the unitarity bounds shown in Eq.~\ref{Eq: amp} as well as the vacuum stability requirement from Eq.~\ref{Eq: tachyon}.}
\end{figure}

\noindent
At large $s$, only the scalar four-point interactions contribute to the scattering amplitudes. Here, we can work in the s-wave if we consider the following set of pairs
\begin{align}
\bigl(hh, \,\rho\rho, \,h\rho \bigr).
\end{align}
In the interaction eigenstate basis, $h\rho$ does not scatter against the first two states so we can consider the $2\times 2$ scattering matrix of the states $hh$ and $\rho\rho$. In the s-wave, this matrix is
\begin{align}
  \mathcal{T} &= \frac{1}{16\pi} \begin{pmatrix}
        6\lambda_1 & \lambda_3\\
        \lambda_3 & 6\lambda_2
    \end{pmatrix}.
\end{align}
Requiring  $\left|\mathcal{T}_{ii}\right| < \frac{1}{2}$ as before leads to
\begin{align}
    3 \left(\lambda_1 + \lambda_2\right) \pm \sqrt{ 9\left(\lambda_1 - \lambda_2\right)^2 + \lambda_3^2} < 8\pi.
\end{align}
The maximum allowed values for $|\lambda_3|$ are shown in Fig.~\ref{Fig: maxl3} in function of $\lambda_1$ and $\lambda_2$. Here, we also require that the vacuum be stable. This requirement provides an upper bound on $\lambda_3$ of the form
\begin{align}
|\lambda_3| < \frac{1}{2}\sqrt{\lambda_1\lambda_2}.
\label{Eq: tachyon}
\end{align}
The highest upper bounds on $\lambda_1$ and $\lambda_2$ are
\begin{align}
\lambda_1, \lambda_2 \lesssim \frac{4\pi}{3}.
\end{align}

\section{Constraints on the Dark Matter Sector}
\label{Sec: sec5}
\subsection{Relic Density}
\label{SubSec: relic}
\begin{figure}
\begin{comment}
\begin{align*}
\begin{gathered}
\begin{fmffile}{DMDMZZ}
\begin{fmfgraph*}(80,80)
\fmfleft{i,j}
\fmfright{o1,o2}
\fmf{fermion,label=$\chi$,l.side=left}{i,v1}
\fmf{fermion,label=$\chi$,l.side=left}{v1,v2}
\fmf{fermion,label=$\bar\chi$,l.side=left}{v2,j}
\fmf{photon,label=$Z'$,l.side=left}{v1,o1}
\fmf{photon,label=$Z'$}{v2,o2}
\end{fmfgraph*}
\end{fmffile} 
\end{gathered}\; \propto g'^4\quad\quad
\begin{gathered}
\begin{fmffile}{DMDMZH}
\begin{fmfgraph*}(80,80)
\fmfleft{i,j}
\fmfright{o1,o2}
\fmf{fermion,label=$\chi$,l.side=left}{i,v1}
\fmf{fermion,label=$\chi$,l.side=left}{v1,v2}
\fmf{fermion,label=$\bar\chi$,l.side=left}{v2,j}
\fmf{photon,label=$Z'$,l.side=left}{v1,o1}
\fmf{dashes,label=$H$}{v2,o2}
\end{fmfgraph*}
\end{fmffile} 
\end{gathered}\; \propto g'^2\lambda_\chi^2\quad\quad
\begin{gathered}
\begin{fmffile}{DMDMHH}
\begin{fmfgraph*}(80,80)
\fmfleft{i,j}
\fmfright{o1,o2}
\fmf{fermion,label=$\chi$,l.side=left}{i,v1}
\fmf{fermion,label=$\chi$,l.side=left}{v1,v2}
\fmf{fermion,label=$\bar\chi$,l.side=left}{v2,j}
\fmf{dashes,label=$H$,l.side=left}{v1,o1}
\fmf{dashes,label=$H$}{v2,o2}
\end{fmfgraph*}
\end{fmffile} 
\end{gathered}\; \propto\lambda_\chi^4\\
\end{align*}
\end{comment}
\includegraphics[width=0.85\linewidth,natwidth=140,natheight=34]{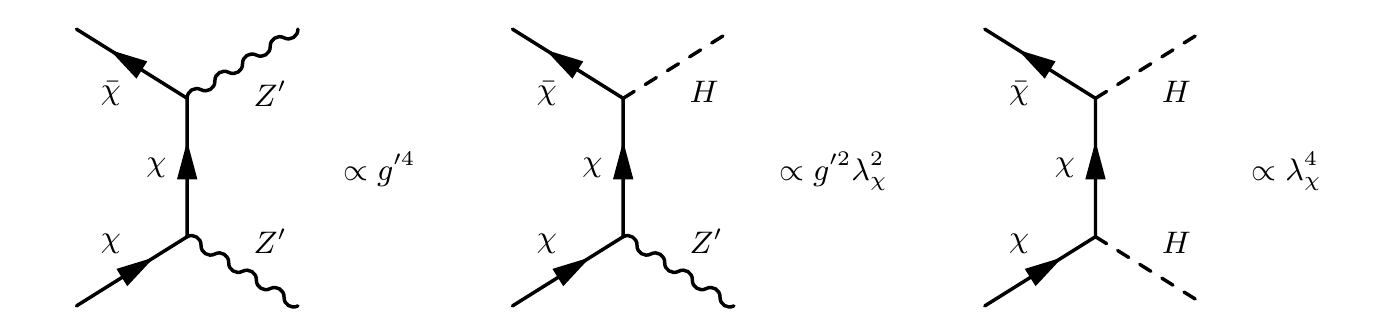}
\caption{\label{Fig: tchannel} t-channel annihilation diagrams for Dark Matter in the $Z'$ model with the corresponding coupling factors.}
\end{figure}

\begin{figure}
\begin{comment}
\begin{align*}
\begin{gathered}
\begin{fmffile}{DM4}
\begin{fmfgraph*}(80,80)
\fmfleft{i,j}
\fmfright{o1,o2,o3,o4}
\fmf{fermion,label=$\chi$,l.side=left}{i,v1}
\fmf{fermion,label=$\chi$,l.side=left}{v1,v2}
\fmf{fermion,label=$\bar\chi$,l.side=left}{v2,j}
\fmf{photon,label=$Z'$,l.side=left}{v1,v3}
\fmf{photon,label=$Z'$}{v2,v4}
\fmf{fermion}{v3,o1}
\fmf{fermion}{v3,o2}
\fmf{fermion}{v4,o3}
\fmf{fermion}{v4,o4}
\end{fmfgraph*}
\end{fmffile} 
\end{gathered}
\end{align*}
\end{comment}
\includegraphics[width=0.2\linewidth,natwidth=32,natheight=34]{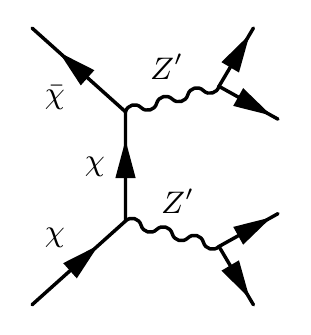}
\caption{\label{Fig: 4body} Example of a $2\rightarrow 4$ Dark Matter annihilation process, via two on-shell $Z'$ bosons. Similar processes with intermediate heavy Higgs bosons are also possible. Each of the intermediate particles decays to two SM particles.}
\end{figure}

\noindent
In the model studied, DM has a large number of possible annihilation channels. We can distinguish two sectors:
\newline
\newline
\paragraph{The s-channel sector:} In this region of parameter space, DM is either lighter than both the dark Higgs and the $Z'$ or close to an s-channel resonance. t-channel annihilation to a dark Higgs or a $Z'$ is then kinematically forbidden in the first case and largely subdominant in the second case. If the $Z'$ boson is not too much heavier than DM and DM is outside the Higgs funnel regions, DM will annihilate dominantly to SM fermions through s-channel $Z'$ exchange. If DM is much heavier than the SM fermions, annihilation occurs dominantly through p-wave. Writing the annihilation cross section times velocity as
\begin{align}
\sigma_\mathrm{ann} v \sim a + b v^2
\end{align}
the p-wave term is approximately
\begin{align}
b \sim g'^2 \tan^2\chi \frac{N_c}{6\pi} \frac{m_\chi^2}{(4 m_\chi^2 - m_{Z'}^2)^2 + \Gamma_{Z'}^2}\left((g^\mathrm{axial}_{Zf\bar f})^2 - (g^\mathrm{vector}_{Zf\bar f})^2\right)
\end{align}
The reason why this annihilation channel dominates for heavy DM is that the couplings associated to the diagram are kinetic mixing suppressed and not mass mixing suppressed. Therefore, they do not vanish when the $Z'$ becomes heavy. Away from the s-channel resonance, the relic density constraint will provide a lower bound on $g'$, which --for heavy DM-- will be in tension with the unitarity bound derived in Sec.~\ref{Sec: sec4}. Close to the s-channel resonance, low values for $g'$ will still be allowed and the bounds on the DM mass will become much looser.  If the $Z'$ is much heavier than DM and the dark Higgs, it can be integrated out and the model reduces to a pure Higgs portal, which has been described in detail in \cite{Walker:2013hka}.
\newline
\newline
\paragraph{The t-channel sector:} As soon as t-channel annihilation to hidden sector particles is kinematically allowed, it becomes the dominant annihilation mode for DM if the latter is away from funnel regions. t-channel annihilation to the hidden sector opens up when the DM is heavier than either the dark Higgs or the dark gauge boson. The corresponding diagrams are shown in Fig.~\ref{Fig: tchannel}. These channels in fact correspond to $2\rightarrow 4$ processes where DM annihilates to four SM particles via on-shell hidden sector particles, as shown in Fig.~\ref{Fig: 4body}. The corresponding cross section can be written as
\begin{align}
\sigma_{\mathrm{ann}} &= \sigma(\chi\bar \chi \rightarrow \mathrm{hidden}_1 + \mathrm{hidden}_2) \,\mathrm{Br}(\mathrm{hidden}_1\rightarrow \mathrm{SM}_1 + \mathrm{SM}_2)\, \mathrm{Br}(\mathrm{hidden}_2\rightarrow \mathrm{SM}_3 + \mathrm{SM}_4).
\end{align}
Generically, the branching ratios for hidden sector decays to SM are very close to $1$. In our study, we assume them to always be $1$, which leads to conservative estimates for the final upper bounds. The annihilation cross section in the t-channel region is then
\begin{align}
\sigma_{\mathrm{ann}} &= \sigma(\chi\bar \chi \rightarrow \mathrm{hidden}_1 + \mathrm{hidden}_2).
\end{align}
Fig.~\ref{Fig: tchannel} shows the coupling products associated to each annihilation diagram. Depending on what diagram dominates, the relic density requirement will then set lower bounds on either $\lambda_\chi$, $g'$ or $\sqrt{\lambda_\chi g'}$. 

\subsection{Direct Detection Bounds}
\begin{figure}
\begin{comment}
\begin{align*}
\begin{gathered}
\begin{fmffile}{directZ}
\begin{fmfgraph*}(80,80)
\fmfleft{i,j}
\fmfright{o1,o2}
\fmf{fermion,label=$u,,d$,l.side=left}{i,v1}
\fmf{fermion,label=$\bar u,,\bar d$,l.side=left}{v1,o1}
\fmf{fermion,label=$\bar\chi$,l.side=left}{v2,j}
\fmf{fermion,label=$\chi$,l.side=left}{o2,v2}
\fmf{photon,label=$Z'$}{v1,v2}
\end{fmfgraph*}
\end{fmffile}
\end{gathered}\quad\quad
\begin{gathered}
\begin{fmffile}{directH}
\begin{fmfgraph*}(80,80)
\fmfleft{i,j}
\fmfright{o1,o2}
\fmf{fermion,label=$u,,d$,l.side=left}{i,v1}
\fmf{fermion,label=$\bar u,,\bar d$,l.side=left}{v1,o1}
\fmf{fermion,label=$\bar\chi$,l.side=left}{v2,j}
\fmf{fermion,label=$\chi$,l.side=left}{o2,v2}
\fmf{dashes,label=$H$}{v1,v2}
\end{fmfgraph*}
\end{fmffile}
\end{gathered}
\end{align*}
\end{comment}
\includegraphics[width=0.5\linewidth,natwidth=66,natheight=36]{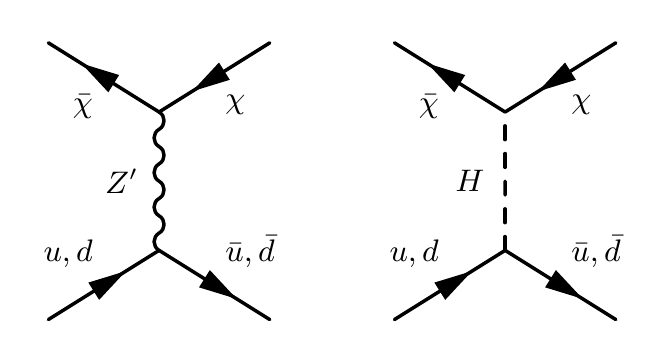}
\caption{\label{Fig: DD} Processes contributing dominantly to the DM direct detection cross section.}
\end{figure}

\noindent
Since both the dark Higgs and the $Z'$ mix with their SM counterpart, DM spin-independent scattering with nuclei is allowed at tree-level. The scattering occurs in the t-channel through the diagrams shown in Fig.~\ref{Fig: DD}. The diagrams associated to a Higgs exchange, however, are mass mixing suppressed and proportional to the small Higgs Yukawa couplings to the $u$ and $d$ quarks. Diagrams with an intermediate $Z'$ are suppressed only by the kinetic mixing. However, the associated spin-independent cross section is velocity suppressed. The spin-independent direct detection (DD) cross section for our model is then expected to be extremely small and the current and future DM DD experiments are not expected to set any significant bounds on the DM mass. Since the current spin-dependent DD bounds are very loose, we do not expect them to give any additional constraint in the heavy DM sector.

\section{Results}
\label{Sec: sec6}
\begin{figure}
\centering
\includegraphics[width=0.55\linewidth,natwidth=200,natheight=200]{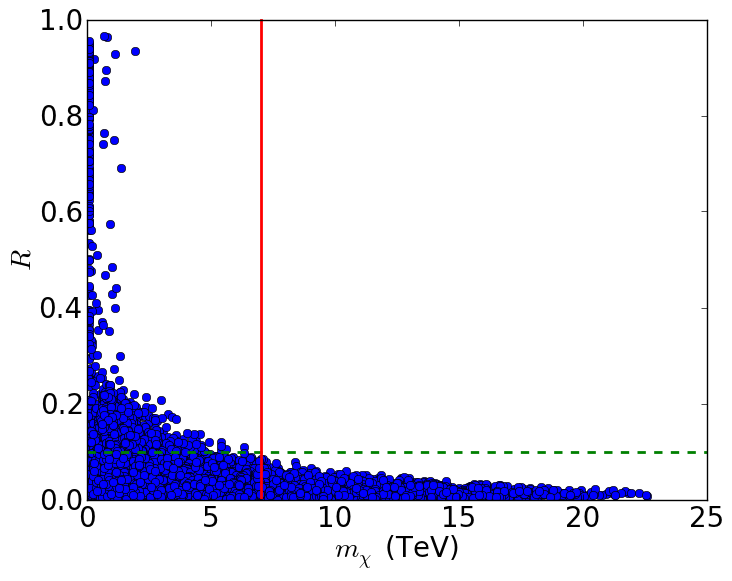}
\caption{\label{Fig: RmixedS} $R$ vs. $m_\chi$ for the points that survive the EWPT, unitarity and relic density cuts and annihilate dominantly through s-channel diagrams. The red line is at $7\mathrm{TeV}$.} % The Higgs sector is constrained to be perturbative.}
\end{figure}

\begin{figure}
\centering
\includegraphics[width=0.55\linewidth,natwidth=200,natheight=200]{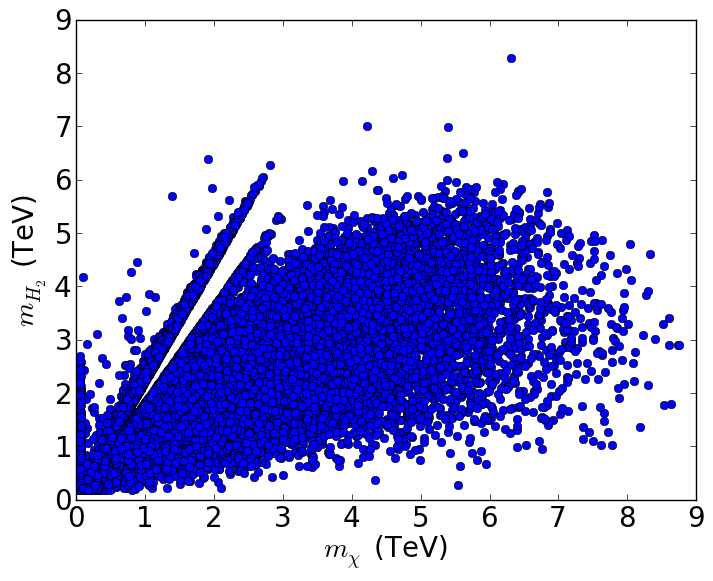}
\caption{\label{Fig: MHMixed} $m_{H_2}$ vs. $m_\chi$ for the points that survive the EWPT, unitarity and relic density cuts for $R > 0.1$. The region around $m_{H_2} \sim 2 m_\chi$ that is sharply cut corresponds to the removed funnel region ($R < 0.1$).}
\end{figure}

\begin{figure}
\centering
\includegraphics[width=0.55\linewidth,natwidth=200,natheight=200]{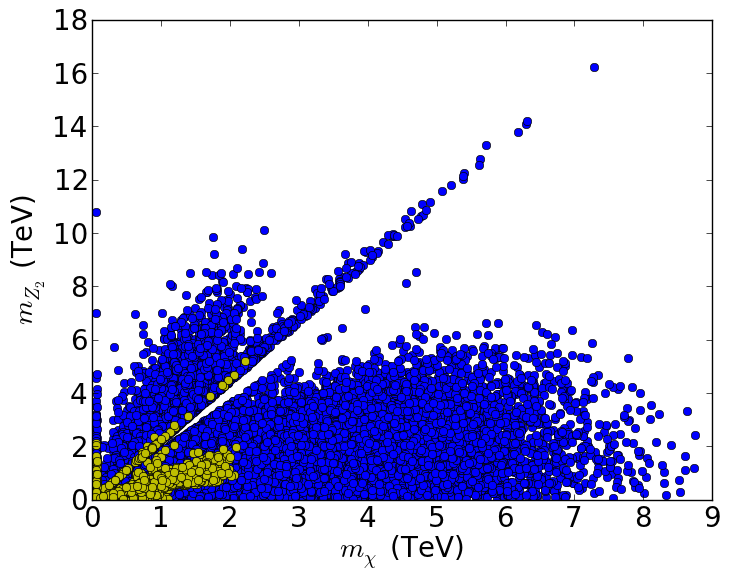}
\caption{\label{Fig: MZMixed} $m_{Z_2}$ vs. $m_\chi$ for the points that survive the EWPT, unitarity and relic density cuts for $R > 0.1$. The region around $m_{H_2} \sim 2 m_\chi$ that is sharply cut corresponds to the removed funnel region ($R < 0.1$). The mixed Higgs-gauge points are shown in blue while the ``pure gauge" points described in Eq.~\ref{Eq: puregauge} are shown in yellow.}
\end{figure}

\noindent
We perform a uniform scan over the set of parameters shown in Eq.~\ref{Eq: parameters}. We fix the coupling $\lambda_1$ by requiring the lightest Higgs mass to be 125 $\mathrm{GeV}$ at tree-level. which reduces the number of parameters to six. 
%We consider two cases
%\begin{itemize}
%\item The Higgs sector is perturbative. In this case, the quartic couplings $\lambda_{1,2,3}$ are constrained by unitarity. The dark Higgs boson $h_2$ then cannot be too heavy and the DM annihilation cross section receives contributions from diagrams with either Higgs or $Z$ bosons.
%\item The Higgs sector is non perturbative. In particular, we consider the case where the dominant DM annihilation channels involve only $Z_2$.
%\end{itemize}
We eliminate points that do not satisfy the EWPT, unitarity and relic abundance constraints detailed in Sec.~\ref{Sec: sec3}, \ref{Sec: sec4}, \ref{Sec: sec5} respectively. We generate the $U(1)_D$ model described in Sec.~\ref{Sec: sec3} using FeynRules \cite{Alloul:2013bka} and compute the DM relic density using MicrOmegas \cite{Belanger:2013oya} interfaced with CalcHEP \cite{Belyaev:2012qa}. We allow the DM, $\chi$, to be a subdominant component of the measured relic abundance, 
\begin{align}
\Omega h^2 \le 0.1199 + 0.0027,
\end{align}
and therefore require the relic density to be less than its Planck value \cite{Ade:2013lta} plus $3\sigma$.
%\subsection{Results with a perturbative Higgs sector}
\newline
\newline
We scan over the following regions of parameter space,
\begin{align}
g' &\in \left[0, 4\right]\\
\lambda_{2} &\in \left[0, 7\right]\\
\lambda_3 &\in \left[-7, 7\right]\\
\lambda_\chi &\in \left[0, 7\right]\\
\sin\chi &\in \left[10^{-3}, 1\right]\\
u &\in \left[0, 50~\mathrm{TeV}\right].
\label{Eq: scanbounds}
\end{align}
We apply the EWPT, unitarity and relic density constraints described in Sec.~\ref{Sec: sec3}, \ref{Sec: sec4} and \ref{Sec: sec5}. The bounds associated to these different constraints are expected to be significantly loosened near s-channel resonances. In order to best identify these ``funnel regions", we define a \emph{fine tuning factor} $R$, as
\begin{align}
R = \mathrm{min}\left( \left|\frac{2 m_\chi - m_{H_2}}{m_{H_2}}\right|,\left|\frac{2 m_\chi - m_{Z_2}}{m_{Z_2}}\right|\right) .
\label{Eq: R}
\end{align}
If a parameter point is close to an s-channel resonance, $R$ will be small. The bounds quoted in this paper hold for $R > 0.1$. Since the low-$R$ funnel regions are particularly narrow, the bounds derived in this work can be considered generic in the context of a $U(1)_D$ model.
\newline
\newline
\begin{figure}
\centering
\includegraphics[width=0.55\linewidth,natwidth=200,natheight=200]{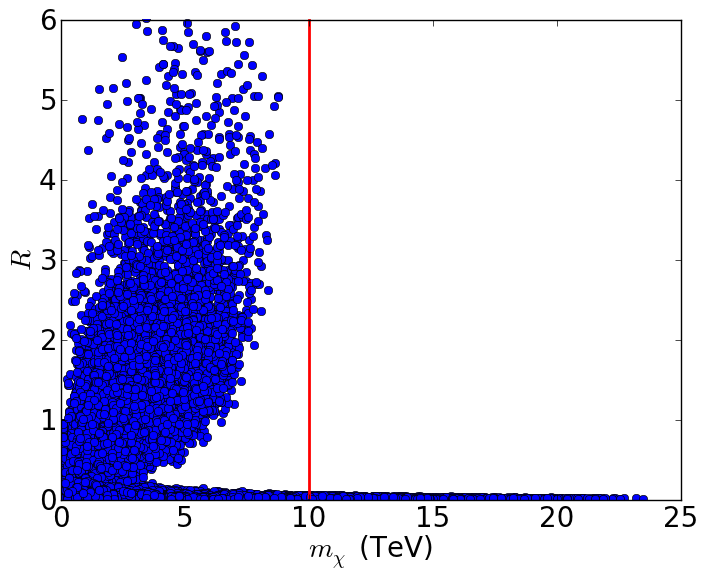}
\caption{\label{Fig: Rmixed} $R$ vs. $m_\chi$ for the points that survive the EWPT, unitarity and relic density cuts. The red line is at $10~\mathrm{TeV}$. The low $R$ funnel region as well as the large $R$ $t$-channel region, both described in the main text, can be clearly distinguished.}
\end{figure}
\begin{figure}
\centering
\includegraphics[width=0.55\linewidth,natwidth=200,natheight=200]{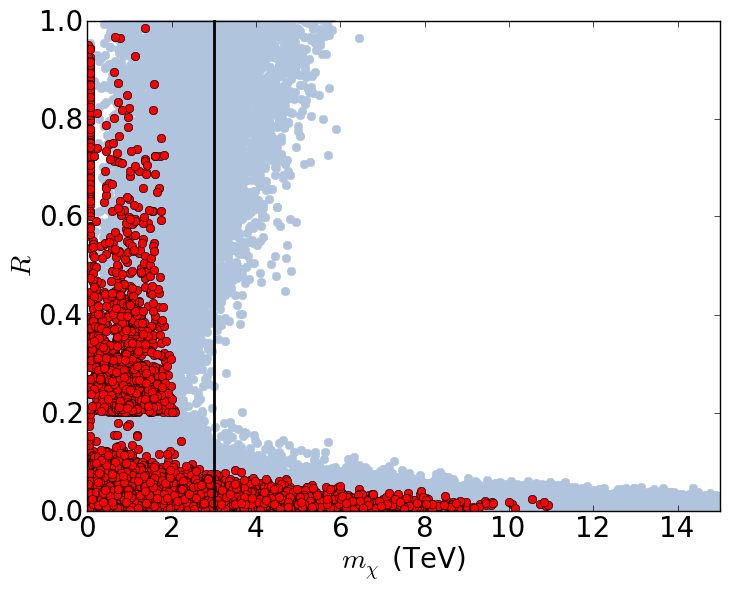}
\caption{\label{Fig: Rgauge} $R$ vs. $m_\chi$ for the points that survive the EWPT, unitarity and relic density cuts in the ``pure gauge" scenario described in Eq.~\ref{Eq: puregauge} (in red). The black line at $3~\mathrm{TeV}$ corresponds to the $R < 10\%$ upper bound on the DM mass for these ``pure gauge" points. The sharp feature around $R = 20\%$ is due to the first condition in Eq.~\ref{Eq: puregauge}. The low $R$ funnel region as well as the large $R$ $t$-channel region, both described in the main text, can be clearly distinguished. The points corresponding to the mixed Higgs-gauge portal scenario shown in Fig.~\ref{Fig: Rmixed} are shown in light blue.}
\end{figure}
\newline
\newline
Fig.~\ref{Fig: Rmixed} shows the points that survive all the EWPT, unitarity and relic density constraints in the $R$ vs. $m_\chi$ plane. In this plane, the surviving points can be grouped into two distinct regions. The first region, also shown isolatedly in  Fig.~\ref{Fig: RmixedS} corresponds to the $s$-channel region described in Sec.~\ref{Sec: sec5}. In this region, DM annihilates dominantly through $s$-channel processes. As can be seen in Fig.~\ref{Fig: RmixedS}, beyond about $2\mathrm{TeV}$, only points close to $s$-channel resonances can satisfy the relic abundance requirement and the allowed values for $R$ become lower as DM gets heavier. For DM masses larger than about $7~\mathrm{ TeV}$ efficient annihilation can occur only for $R < 10\%$.  The second region in Fig.~\ref{Fig: Rmixed} corresponds to the $t$-channel region introduced in Sec.~\ref{Sec: sec5}. Here, DM annihilation is dominated by $t$-channel annihilations to dark Higgs and/or gauge bosons. For $R < 1$, only a few of these processes are kinematically allowed. As can be seen in Fig.~\ref{Fig: Rmixed}, $R$ then needs to increase with the DM mass in order to open more annihilation channels. For $R\gsim 1$, all the $t$-channel annihilation modes are allowed and the only way to increase the DM annihilation cross section is to increase the DM couplings. When these couplings reach their unitarity bounds, the DM mass reaches its maximum, which corresponds to the vertical cutoff for $R\gsim 1$ on Fig.~\ref{Fig: Rmixed}. This cutoff corresponds to a DM mass of $10~\mathrm{TeV}$.
%and correspond to the $s$ and $t$-channel regions described in Sec.~\ref{Sec: sec5}. In the s-channel region, also shown alone in Fig.~\ref{Fig: RmixedS}, the allowed points of parameter space have to get closer and closer to the funnel regions as DM gets heavier. The t-channel region, on the contrary, exhibits a hard cutoff around about $10~\mathrm{TeV}$. This cutoff corresponds to the $g'$ and/or $\lambda_\chi$ couplings reaching the maximum values allowed by unitarity. Here, instead of decreasing as in the s-channel region, $R$ increases smoothly. This increase in $R$ can be explained by the fact that, as DM gets heavier, more and more annihilation channels need to open up in order for DM to annihilate efficiently enough. For $R \gsim 1$, in particular, all the possible t-channel annihilation modes are kinematically allowed and increasing $R$ no longer improves the bound on the DM mass.
\newline
\newline
Fig.~\ref{Fig: Rgauge} shows the points that survive all the cuts and for which Dark Matter annihilate dominantly through modes involving only dark gauge bosons. We identify this ``pure gauge" sector using the following criteria
\begin{align}
&\frac{|2m_\chi - m_{H_2}|}{m_{H_2}} > 20\%\\
&m_{\chi} < \frac{m_{H_2} + m_{Z_2}}{2}\\
&m_{Z_2} < m_{H_2}.
\label{Eq: puregauge}
\end{align}
The first requirement cuts away the dark Higgs funnel region while the second and third requirements forbid t-channel DM annihilation to final states involving a dark Higgs boson. For $R > 0.1$, the bound on the DM mass in this ``pure gauge" sector is of about $3~\mathrm{TeV}$.
\newline
\newline
Figs~\ref{Fig: MHMixed} and  \ref{Fig: MZMixed} show the non-resonant $R > 0.1$ points that survive all the EWPT, unitarity and relic density constraints in the $(m_\chi, m_{H_2})$ plane and the $(m_\chi, m_{Z_2})$ plane, respectively. The blue points correspond to the mixed Higgs-gauge portal scenario while the yellow points on Fig.~\ref{Fig: MZMixed} correspond to the ``pure gauge scenario". The bound on the dark Higgs mass is of about $9~\mathrm{TeV}$. The bound on the $Z_2$ mass is of about $16~\mathrm{TeV}$ for $R > 0.1$. However, except for a few points close to the funnel regions, most points of the parameter space correspond to $Z_2$ masses below $10~\mathrm{TeV}$. If we consider only the ``pure gauge" points, the bound on the dark gauge boson mass goes down to $6~\mathrm{TeV}$. For perturbative DM and Higgs sectors, unitarity and relic density constraints then allow to set a bound of about $10~\mathrm{TeV}$ on the masses of the DM and the dark Higgs and gauge bosons.
\begin{figure}
\centering
\includegraphics[width=0.55\linewidth,natwidth=200,natheight=200]{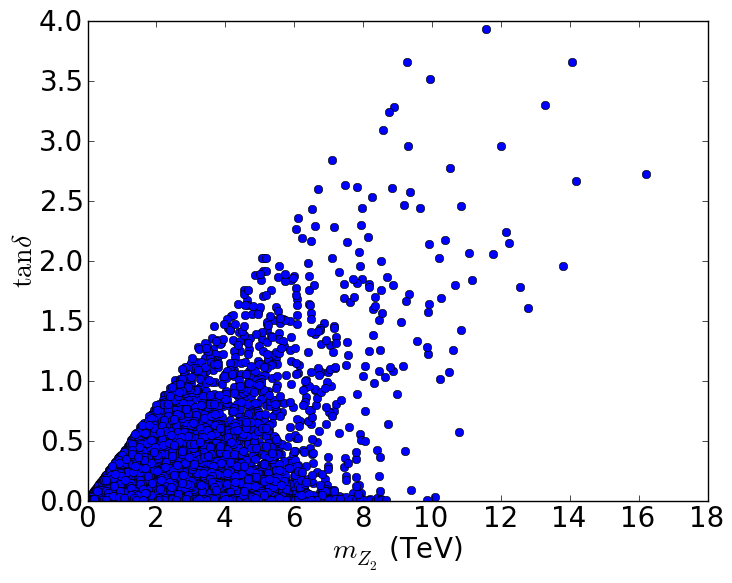}
\caption{\label{Fig:MZpvgeff} $m_{Z_2}$ vs. $\tan\delta$ for the points that survive the EWPT, unitarity and relic density cuts.  In addition, we require $R < 0.1$.}
\end{figure}
\newline
\newline
%\subsection{Impact of Future Experiments}
%\noindent
The increasing sensitivity of accelerator, direct and indirect searches for dark matter provides an opportunity to definitively probe the gauge portal.  %we give a basic discussion of the potential of each and postpone the bulk of our analysis to future work.  
In Figure~\ref{Fig:MZpvgeff}, we plot $m_{Z_2}$ versus $\tan\delta$.  As a reminder, because of equation~\ref{eq:ffZprimecoupling}, the $\bar{f} f \to Z' \to \bar{f} f$ cross section scales as
\begin{equation}
\sigma_{\bar{f} f \to Z' \to \bar{f} f} \equiv \tan^4\delta \,\,\sigma_{\mathrm{SM}\,\,\bar{f} f \to Z \to \bar{f} f}.
\end{equation}
A few of the points in Figure~\ref{Fig:MZpvgeff} for $\tan\delta > 1$ may lead to a greater sensitivity for accelerator and indirect searches in comparison to sequential SM $Z'$ models.  The points for $\tan\delta < 1$ may be problematic.   Moreover as described above, our model enhances the effective operators
\begin{align}
\mathcal{O}_1 = {1 \over \Lambda_1^2} \bar{q} \gamma^\mu \gamma_5 q\, \bar{\chi}\gamma_\mu\chi && \mathcal{O}_2 = {1 \over \Lambda_3^2} \bar{q} \gamma^\mu \gamma_5 q\, \bar{\chi}\gamma_\mu \gamma_5 \chi
\end{align}
which correspond to a DM-nucleon scattering cross section that is either velocity suppressed or spin-dependent.  Thus, the spin-independent dark matter-nucleon cross section, to which the current experiments are highly sensitive, is small.  A different model with a non-trivial operator
\begin{align}
\mathcal{O}_3 = {1 \over \Lambda_2^2} \bar{q} \gamma^\mu q\, \bar{\chi}\gamma_\mu\chi
\end{align}
could give a significant signal in direct detection experiments.  We postpone a thorough examination of these issues for future work.
\newline
\newline

\section{Conclusion}
\label{Sec: conclusion}
\noindent
In this paper, we showed that combining perturbative unitarity and relic abundance constraints on dark matter in a generic gauge portal model allows upper bounds to be set on the dark matter, dark $Z'$ and dark Higgs masses.  % for most of the parameter space. 
%Even with minimal assumptions about the $Z'$ model and the DM properties, 
The bounds derived here are significantly improved with respect to the Griest and Kamionkowski bound \cite{Griest:1989wd}. The bounds shown in this paper are valid for all the model parameter space except in the narrow dark Higgs and $Z$ funnel regions. 
\newline
\newline
Finding the next scale of new physics is a crucial question for current and future experiments.  Although naturalness provides strong arguments in favor of new physics around the $\mathrm{TeV}$ scale, the existence of dark matter is one of the most compelling reason for new physics. Using fundamental principles in concert with experimental measurements to better constrain the DM sector provides a new avenue for cornering new physics at future colliders and DM experiments. For the model of dark matter explored here the most sensitive probes are future collider searches for new $Z'$ bosons. We find that there is a finite mass window within which dark matter and its associated new particles can appear, requiring that the $Z'$ be lighter than $6~\rm{TeV}$ if it is dominantly responsible for dark matter annihilations. This is a very promising region of parameter space for future high-energy colliders.


\begin{thebibliography}{000}


\bibitem{Kowalski:2008ez} 
  M.~Kowalski {\it et al.}  [Supernova Cosmology Project Collaboration],
  %``Improved Cosmological Constraints from New, Old and Combined Supernova Datasets,''
  Astrophys.\ J.\  {\bf 686}, 749 (2008)
  [arXiv:0804.4142 [astro-ph]].
  %%CITATION = ARXIV:0804.4142;%%
  %802 citations counted in INSPIRE as of 31 Aug 2013

 \bibitem{Ahn:2013gms} 
  C.~P.~Ahn {\it et al.}  [SDSS Collaboration],
  %``The Tenth Data Release of the Sloan Digital Sky Survey: First Spectroscopic Data from the SDSS-III Apache Point Observatory Galactic Evolution Experiment,''
  arXiv:1307.7735 [astro-ph.IM].
  %%CITATION = ARXIV:1307.7735;%%
  %1 citations counted in INSPIRE as of 31 Aug 2013
  
  \bibitem{Beringer:1900zz} 
  J.~Beringer {\it et al.}  [Particle Data Group Collaboration],
  %``Review of Particle Physics (RPP),''
  Phys.\ Rev.\ D {\bf 86}, 010001 (2012).
  %%CITATION = PHRVA,D86,010001;%%
   
\bibitem{Ade:2013lta} 
  P.~A.~R.~Ade {\it et al.}  [ Planck Collaboration],
  %``Planck 2013 results. XVI. Cosmological parameters,''
  arXiv:1303.5076 [astro-ph.CO].
  %%CITATION = ARXIV:1303.5076;%%
  %13 citations counted in INSPIRE as of 28 Mar 2013
  
  \bibitem{Bertone:2004pz} 
  G.~Bertone, D.~Hooper and J.~Silk,
  %``Particle dark matter: Evidence, candidates and constraints,''
  Phys.\ Rept.\  {\bf 405}, 279 (2005)
  [hep-ph/0404175].
  %%CITATION = HEP-PH/0404175;%%
  %1544 citations counted in INSPIRE as of 15 Sep 2013
   
\bibitem{Aad:2014tda} 
  G.~Aad {\it et al.}  [ ATLAS Collaboration],
  %``Search for new phenomena in events with a photon and missing transverse momentum in $pp$ collisions at $\sqrt{s}=8$ TeV with the ATLAS detector,''
  arXiv:1411.1559 [hep-ex].
  %%CITATION = ARXIV:1411.1559;%%
     
\bibitem{ATLAS:2014wra} 
  G.~Aad {\it et al.}  [ATLAS Collaboration],
  %``Search for new particles in events with one lepton and missing transverse momentum in $pp$ collisions at $\sqrt{s}$ = 8 TeV with the ATLAS detector,''
  JHEP {\bf 1409}, 037 (2014)
  [arXiv:1407.7494 [hep-ex]].
  %%CITATION = ARXIV:1407.7494;%%
  %11 citations counted in INSPIRE as of 14 Dec 2014
   
\bibitem{Aad:2013oja} 
  G.~Aad {\it et al.}  [ATLAS Collaboration],
  %``Search for dark matter in events with a hadronically decaying W or Z boson and missing transverse momentum in $pp$ collisions at $\sqrt{s} =$ 8 TeV with the ATLAS detector,''
  Phys.\ Rev.\ Lett.\  {\bf 112}, no. 4, 041802 (2014)
  [arXiv:1309.4017 [hep-ex]].
  %%CITATION = ARXIV:1309.4017;
  %%56 citations counted in INSPIRE as of 14 Dec 2014
  
  \bibitem{Aad:2014vka} 
  G.~Aad {\it et al.}  [ATLAS Collaboration],
  %``Search for dark matter in events with a Z boson and missing transverse momentum in pp collisions at $\sqrt{s}$=8 TeV with the ATLAS detector,''
  Phys.\ Rev.\ D {\bf 90}, no. 1, 012004 (2014)
  [arXiv:1404.0051 [hep-ex]].
  %%CITATION = ARXIV:1404.0051;%%
  %26 citations counted in INSPIRE as of 14 Dec 2014
  
  \bibitem{Aad:2014kra} 
  G.~Aad {\it et al.}  [ATLAS Collaboration],
  %``Search for top squark pair production in final states with one isolated lepton, jets, and missing transverse momentum in $\sqrt s =$8 TeV $pp$ collisions with the ATLAS detector,''
  JHEP {\bf 1411}, 118 (2014)
  [arXiv:1407.0583 [hep-ex]].
  %%CITATION = ARXIV:1407.0583;%%
  %34 citations counted in INSPIRE as of 14 Dec 2014
  
  \bibitem{Chatrchyan:2014lfa} 
  S.~Chatrchyan {\it et al.}  [CMS Collaboration],
  %``Search for new physics in the multijet and missing transverse momentum final state in proton-proton collisions at $\sqrt{s}$= 8 TeV,''
  JHEP {\bf 1406}, 055 (2014)
  [arXiv:1402.4770 [hep-ex]].
  %%CITATION = ARXIV:1402.4770;%%
  %74 citations counted in INSPIRE as of 14 Dec 2014
  
 \bibitem{Khachatryan:2014tva} 
  V.~Khachatryan {\it et al.}  [CMS Collaboration],
  %``Search for physics beyond the standard model in final states with a lepton and missing transverse energy in proton-proton collisions at $\sqrt{s}$ = 8 TeV,''
  arXiv:1408.2745 [hep-ex].
  %%CITATION = ARXIV:1408.2745;%%
  %11 citations counted in INSPIRE as of 14 Dec 2014
  
  \bibitem{Khachatryan:2014rwa} 
  V.~Khachatryan {\it et al.}  [CMS Collaboration],
  %``Search for new phenomena in monophoton final states in proton-proton collisions at sqrt(s) = 8 TeV,''
  arXiv:1410.8812 [hep-ex].
  %%CITATION = ARXIV:1410.8812;%%
   
  \bibitem{Khachatryan:2014rra} 
  V.~Khachatryan {\it et al.}  [CMS Collaboration],
  %``Search for dark matter, extra dimensions, and unparticles in monojet events in proton-proton collisions at $\sqrt{s}$ = 8 TeV,''
  arXiv:1408.3583 [hep-ex].
  %%CITATION = ARXIV:1408.3583;%%
  %28 citations counted in INSPIRE as of 14 Dec 2014
  
  \bibitem{Chatrchyan:2013lya} 
  S.~Chatrchyan {\it et al.}  [CMS Collaboration],
  %``Search for supersymmetry in hadronic final states with missing transverse energy using the variables $\alpha_T$ and b-quark multiplicity in pp collisions at $\sqrt s=8$ TeV,''
  Eur.\ Phys.\ J.\ C {\bf 73}, no. 9, 2568 (2013)
  [arXiv:1303.2985 [hep-ex]].
  %%CITATION = ARXIV:1303.2985;%%
  %121 citations counted in INSPIRE as of 14 Dec 2014
   
   \bibitem{Chatrchyan:2013xna} 
  S.~Chatrchyan {\it et al.}  [CMS Collaboration],
  %``Search for top-squark pair production in the single-lepton final state in pp collisions at $\sqrt{s}$ = 8 TeV,''
  Eur.\ Phys.\ J.\ C {\bf 73}, no. 12, 2677 (2013)
  [arXiv:1308.1586 [hep-ex]].
  %%CITATION = ARXIV:1308.1586;%%
  %108 citations counted in INSPIRE as of 14 Dec 2014
  
  \bibitem{Aaltonen:2012jb} 
  T.~Aaltonen {\it et al.}  [CDF Collaboration],
  %``A Search for dark matter in events with one jet and missing transverse energy in $p\bar{p}$ collisions at $\sqrt{s} = 1.96$ TeV,''
  Phys.\ Rev.\ Lett.\  {\bf 108}, 211804 (2012)
  [arXiv:1203.0742 [hep-ex]].
  %%CITATION = ARXIV:1203.0742;%%
  %61 citations counted in INSPIRE as of 15 Dec 2014
   
   \bibitem{Aaltonen:2013har} 
  T.~Aaltonen {\it et al.}  [CDF Collaboration],
  %``Signature-based search for delayed photons in exclusive photon plus missing transverse energy events from $p \bar p$ collisions with $\sqrt s=1.96?$?TeV,''
  Phys.\ Rev.\ D {\bf 88}, no. 3, 031103 (2013)
  [arXiv:1307.0474 [hep-ex]].
  %%CITATION = ARXIV:1307.0474;%%
  %2 citations counted in INSPIRE as of 15 Dec 2014
  
  \bibitem{CDF:2011ah} 
  T.~Aaltonen {\it et al.}  [CDF Collaboration],
  %``Search for new phenomena in events with two $Z$ bosons and missing transverse momentum in $p\bar{p}$ collisions at $\sqrt{s}=1.96$ TeV,''
  Phys.\ Rev.\ D {\bf 85}, 011104 (2012)
  [arXiv:1112.1577 [hep-ex]].
  %%CITATION = ARXIV:1112.1577;%%
  %3 citations counted in INSPIRE as of 15 Dec 2014
  
  \bibitem{Abachi:1996dc} 
  S.~Abachi {\it et al.}  [D0 Collaboration],
  %``Search for diphoton events with large missing transverse energy in $p\bar{p}$ collisions at $\sqrt{s} = 1.8$ TeV,''
  Phys.\ Rev.\ Lett.\  {\bf 78}, 2070 (1997)
  [hep-ex/9612011].
  %%CITATION = HEP-EX/9612011;%%
  %51 citations counted in INSPIRE as of 15 Dec 2014
  
  \bibitem{Abazov:2012qka} 
  V.~M.~Abazov {\it et al.}  [D0 Collaboration],
  %``Search for $Z\gamma$ events with large missing transverse energy in $p \bar{p}$ collisions at $\sqrt{s}=1.96$ TeV,''
  Phys.\ Rev.\ D {\bf 86}, 071701 (2012)
  [arXiv:1203.5311 [hep-ex]].
  %%CITATION = ARXIV:1203.5311;%%
  %5 citations counted in INSPIRE as of 15 Dec 2014
  
  \bibitem{Abdallah:2003np} 
  J.~Abdallah {\it et al.}  [DELPHI Collaboration],
  %``Photon events with missing energy in e+ e- collisions at s**(1/2) = 130-GeV to 209-GeV,''
  Eur.\ Phys.\ J.\ C {\bf 38}, 395 (2005)
  [hep-ex/0406019].
  %%CITATION = HEP-EX/0406019;%%
  %126 citations counted in INSPIRE as of 15 Dec 2014
  
  \bibitem{Acciarri:1997dq} 
  M.~Acciarri {\it et al.}  [L3 Collaboration],
  %``Single and multiphoton events with missing energy in e+ e- collisions at 161-GeV < s**(1/2) < 172-GeV,''
  Phys.\ Lett.\ B {\bf 415}, 299 (1997).
  %%CITATION = PHLTA,B415,299;%%
  %25 citations counted in INSPIRE as of 15 Dec 2014
  
  \bibitem{Abbiendi:2004gf} 
  G.~Abbiendi {\it et al.}  [OPAL Collaboration],
  %``Multi-photon events with large missing energy in e+ e- collisions at s**(1/2) = 192-GeV - 209-GeV,''
  Phys.\ Lett.\ B {\bf 602}, 167 (2004)
  [hep-ex/0412011].
  %%CITATION = HEP-EX/0412011;%%
  %10 citations counted in INSPIRE as of 15 Dec 2014
   
   \bibitem{Aprile:2012nq} 
  E.~Aprile {\it et al.}  [XENON100 Collaboration],
  %``Dark Matter Results from 225 Live Days of XENON100 Data,''
  Phys.\ Rev.\ Lett.\  {\bf 109}, 181301 (2012).
%  [arXiv:1207.5988 [astro-ph.CO]].
  %%CITATION = ARXIV:1207.5988;%%
  %186 citations counted in INSPIRE as of 11 Apr 2013

  \bibitem{Ahmed:2011gh} 
  Z.~Ahmed {\it et al.}  [CDMS and EDELWEISS Collaborations],
  %``Combined Limits on WIMPs from the CDMS and EDELWEISS Experiments,''
  Phys.\ Rev.\ D {\bf 84}, 011102 (2011).
%  [arXiv:1105.3377 [astro-ph.CO]].
  %%CITATION = ARXIV:1105.3377;%%
  %41 citations counted in INSPIRE as of 11 Apr 2013
  
  %\cite{Akerib:2013tjd}
\bibitem{Akerib:2013tjd} 
  D.~S.~Akerib {\it et al.}  [LUX Collaboration],
  %``First results from the LUX dark matter experiment at the Sanford Underground Research Facility,''
  Phys.\ Rev.\ Lett.\  {\bf 112}, 091303 (2014)
  [arXiv:1310.8214 [astro-ph.CO]].
  %%CITATION = ARXIV:1310.8214;%%
  %386 citations counted in INSPIRE as of 16 Sep 2014

  \bibitem{Baak:2012kk} 
  M.~Baak, M.~Goebel, J.~Haller, A.~Hoecker, D.~Kennedy, R.~Kogler, K.~Moenig and M.~Schott {\it et al.},
  %``The Electroweak Fit of the Standard Model after the Discovery of a New Boson at the LHC,''
  Eur.\ Phys.\ J.\ C {\bf 72}, 2205 (2012).
 % [arXiv:1209.2716 [hep-ph]].
  %%CITATION = ARXIV:1209.2716;%%
  
  \bibitem{devinjoannetim}
R.~Cotta, J.~Hewett, T.~Tait and D.~G.~E.~Walker, to appear.
  
    \bibitem{ALEPH:2005ab} 
  S.~Schael {\it et al.}  [ALEPH and DELPHI and L3 and OPAL and SLD and LEP Electroweak Working Group and SLD Electroweak Group and SLD Heavy Flavour Group Collaborations],
  %``Precision electroweak measurements on the $Z$ resonance,''
  Phys.\ Rept.\  {\bf 427}, 257 (2006)
  [hep-ex/0509008].
  %%CITATION = HEP-EX/0509008;%%
  %849 citations counted in INSPIRE as of 15 Sep 2013 

  \bibitem{Patt:2006fw} 
  B.~Patt and F.~Wilczek,
  %``Higgs-field portal into hidden sectors,''
  hep-ph/0605188.
  %%CITATION = HEP-PH/0605188;%%
  %170 citations counted in INSPIRE as of 23 Aug 2013
  
 \bibitem{Walker:2013hka} 
  D.~G.~E.~Walker,
  %``Unitarity Constraints on Higgs Portals,''
  arXiv:1310.1083 [hep-ph].
  %%CITATION = ARXIV:1310.1083;%%
  %5 citations counted in INSPIRE as of 16 Sep 2014

\bibitem{Betre:2014sra} 
  K.~Betre, S.~E.~Hedri and D.~G.~E.~Walker,
  %``Perturbative Unitarity Constraints on a Supersymmetric Higgs Portal,''
  arXiv:1407.0395 [hep-ph].
  %%CITATION = ARXIV:1407.0395;%%
 
\bibitem{Betre:2014fva} 
  K.~Betre, S.~E.~Hedri and D.~G.~E.~Walker,
  %``Perturbative Unitarity Constraints on the NMSSM Higgs Sector,''
  arXiv:1410.1534 [hep-ph].
  %%CITATION = ARXIV:1410.1534;%%
      
  \bibitem{Griest:1989wd} 
  K.~Griest and M.~Kamionkowski,
  %``Unitarity Limits on the Mass and Radius of Dark Matter Particles,''
  Phys.\ Rev.\ Lett.\  {\bf 64}, 615 (1990).
  %%CITATION = PRLTA,64,615;%%
    
  \bibitem{Shepherd:2009sa} 
  W.~Shepherd, T.~M.~P.~Tait and G.~Zaharijas,
  %``Bound states of weakly interacting dark matter,''
  Phys.\ Rev.\ D {\bf 79}, 055022 (2009)
  [arXiv:0901.2125 [hep-ph]].
  %%CITATION = ARXIV:0901.2125;%%
  %50 citations counted in INSPIRE as of 16 Sep 2014
   
 \bibitem{Profumo:2013yn} 
  S.~Profumo,
  %``TASI 2012 Lectures on Astrophysical Probes of Dark Matter,''
  arXiv:1301.0952 [hep-ph].
  %%CITATION = ARXIV:1301.0952;%%
  %9 citations counted in INSPIRE as of 10 Sep 2014
   
%\cite{Babu:1997st}
\bibitem{Babu:1997st} 
  K.~S.~Babu, C.~F.~Kolda and J.~March-Russell,
  %``Implications of generalized Z - Z-prime mixing,''
  Phys.\ Rev.\ D {\bf 57}, 6788 (1998)
  [hep-ph/9710441].
  %%CITATION = HEP-PH/9710441;%%
  %146 citations counted in INSPIRE as of 15 Dec 2014

%\cite{Wells:2008xg}
\bibitem{Wells:2008xg} 
  J.~D.~Wells,
  %``How to Find a Hidden World at the Large Hadron Collider,''
  In *Kane, Gordon (ed.), Pierce, Aaron (ed.): Perspectives on LHC physics* 283-298
  [arXiv:0803.1243 [hep-ph]].
  %%CITATION = ARXIV:0803.1243;%%
  %20 citations counted in INSPIRE as of 15 Dec 2014

%\cite{Holdom:1990xp}
\bibitem{Holdom:1990xp} 
  B.~Holdom,
  %``Oblique electroweak corrections and an extra gauge boson,''
  Phys.\ Lett.\ B {\bf 259}, 329 (1991).
  %%CITATION = PHLTA,B259,329;%%
  %153 citations counted in INSPIRE as of 15 Dec 2014

     \bibitem{Peskin:1991sw} 
  M.~E.~Peskin and T.~Takeuchi,
  %``Estimation of oblique electroweak corrections,''
  Phys.\ Rev.\ D {\bf 46}, 381 (1992).
  %%CITATION = PHRVA,D46,381;%%
  %1272 citations counted in INSPIRE as of 22 May 2013 
  
 \bibitem{Aydemir:2012nz} 
  U.~Aydemir, M.~M.~Anber and J.~F.~Donoghue,
  %``Self-healing of unitarity in effective field theories and the onset of new physics,''
  Phys.\ Rev.\ D {\bf 86}, 014025 (2012)
  [arXiv:1203.5153 [hep-ph]].
  %%CITATION = ARXIV:1203.5153;%%
  %14 citations counted in INSPIRE as of 16 Dec 2014
 
\bibitem{Schuessler:2007av} 
  A.~Schuessler and D.~Zeppenfeld,
  %``Unitarity constraints on MSSM trilinear couplings,''
  In *Karlsruhe 2007, SUSY 2007* 236-239
  [arXiv:0710.5175 [hep-ph]].
  
   \bibitem{Schuessler:thesis} 
  A.~Schuessler,
%``Unitarit\"{a}ts-Schranken~an~triskalare~Kopplungen im MSSM," 
Diplomarbeit, Institut f\"{u}r Theoretische Physik, Universit\"{a}t Karlsruhe (2005).
%www.itp.kit.edu/diplomatheses.de.shtml (German)

%\cite{Alloul:2013bka}
\bibitem{Alloul:2013bka} 
  A.~Alloul, N.~D.~Christensen, C.~Degrande, C.~Duhr and B.~Fuks,
  %``FeynRules  2.0 - A complete toolbox for tree-level phenomenology,''
  Comput.\ Phys.\ Commun.\  {\bf 185}, 2250 (2014)
  [arXiv:1310.1921 [hep-ph]].
  %%CITATION = ARXIV:1310.1921;%%
  %126 citations counted in INSPIRE as of 17 Dec 2014

%\cite{Belanger:2013oya}
\bibitem{Belanger:2013oya} 
  G.~Belanger, F.~Boudjema, A.~Pukhov and A.~Semenov,
  %``micrOMEGAs_3: A program for calculating dark matter observables,''
  Comput.\ Phys.\ Commun.\  {\bf 185}, 960 (2014)
  [arXiv:1305.0237 [hep-ph]].
  %%CITATION = ARXIV:1305.0237;%%
  %154 citations counted in INSPIRE as of 17 Dec 2014

%\cite{Belyaev:2012qa}
\bibitem{Belyaev:2012qa} 
  A.~Belyaev, N.~D.~Christensen and A.~Pukhov,
  %``CalcHEP 3.4 for collider physics within and beyond the Standard Model,''
  Comput.\ Phys.\ Commun.\  {\bf 184}, 1729 (2013)
  [arXiv:1207.6082 [hep-ph]].
  %%CITATION = ARXIV:1207.6082;%%
  %176 citations counted in INSPIRE as of 17 Dec 2014

    
 
\end{thebibliography}
\end{document}